\documentstyle[prl,aps,twocolumn,floats,graphicx]{revtex}

\newcommand{\position}{tb}
\newcommand{\figA}{
\begin{figure}[\position]
\centerline{\includegraphics{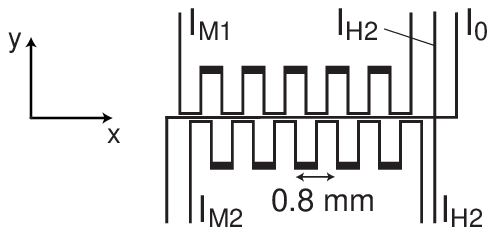}} \caption{Layout of
the lithographic gold wires on the substrate.}
\label{fig-motorLayout}
\end{figure}
}

\newcommand{\figB}{
\begin{figure}[\position]
\centerline{\includegraphics{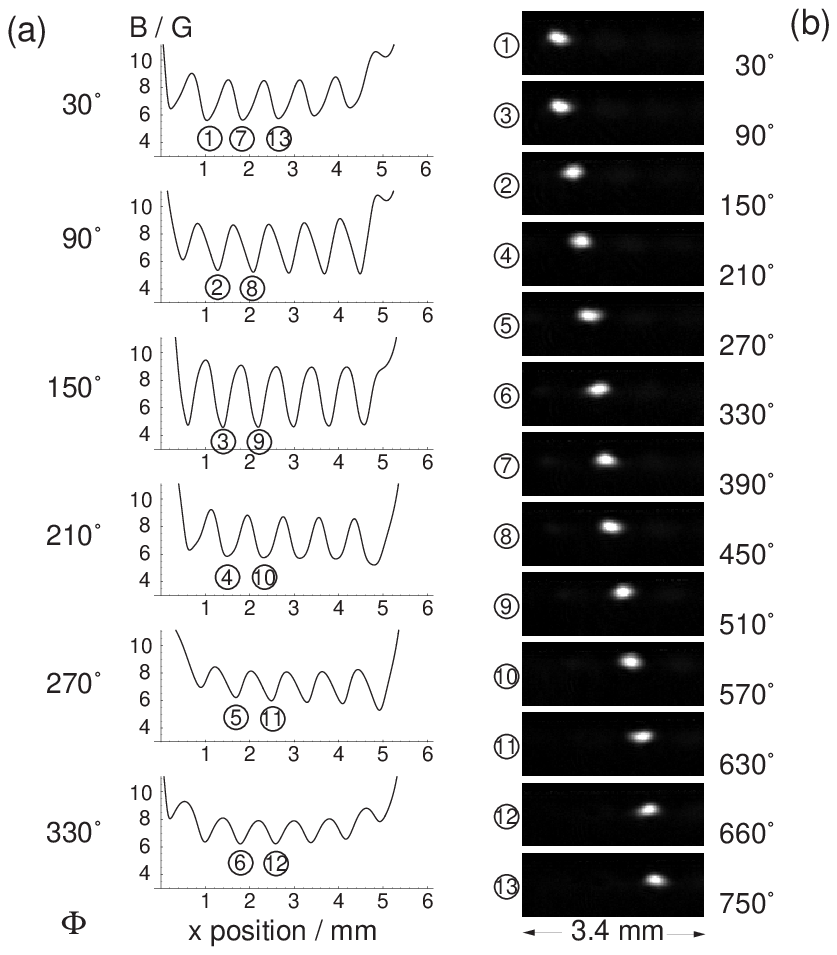}} \caption{(a)
Conveyor belt potential: For each x position the minimum magnetic
field strength in the $\vecy$-$\vecz$ plane is shown. The
potential is created by applying the currents $\I0=2A$,
$(\I{M1},\I{M2})=1\,\mbox{A}\,(\cos\Phi,-\sin\Phi)$ in the wire
configuration of fig.~\ref{fig-motorLayout} and superimposing a
constant bias field
$\vB{0}=7\,\mbox{G}\,\vecx+16\,\mbox{G}\,\vecy$. (b) Absorption
images of an atom cloud transported in this potential. In this
experiment, $\Phi=2\pi\,t/150\,\mbox{ms}$, leading to an average
transport speed of $\overline v=5.3$\,mm$/$s.}
\label{fig-motorComplete}
\end{figure}
}

\newcommand{\figC}{
\begin{figure}[\position]
\centerline{\includegraphics{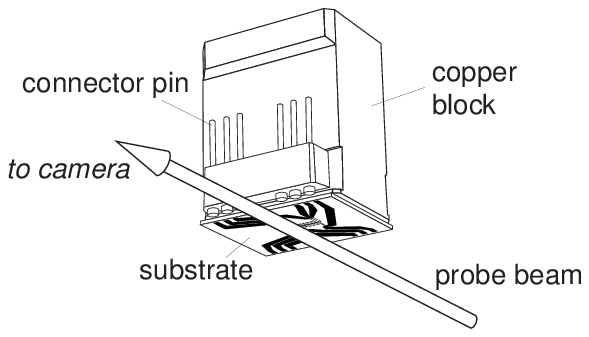}} \caption{Experimental
geometry.} \label{fig-probeBeam}
\end{figure}
}

\newcommand{\figD}{
\begin{figure}[\position]
\centerline{\includegraphics{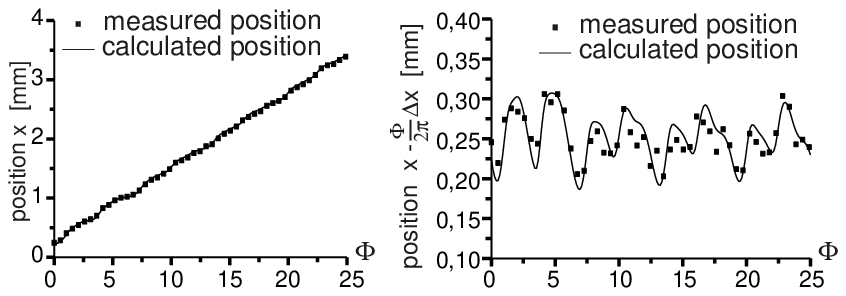}} \caption{(a)
Position of an atom cloud transported in the potential of
fig.~\ref{fig-motorComplete} (a). The center of mass is shown as a
function of the phase angle $\Phi$, which is varied according to
$\Phi=\omega t$ with $\omega=\frac{2\pi}{150\,\text{ms}}$. The
fact that the x position is not strictly linear in $\Phi$ is due
to the particular form of the modulation wires and is well
reproduced in the simulation as can be seen in (b) where the
linear part of $x(\Phi)$ is subtracted.} \label{fig-trajectory}
\end{figure}
}

\newcommand{\figE}{
\begin{figure}[\position]
\centerline{\includegraphics{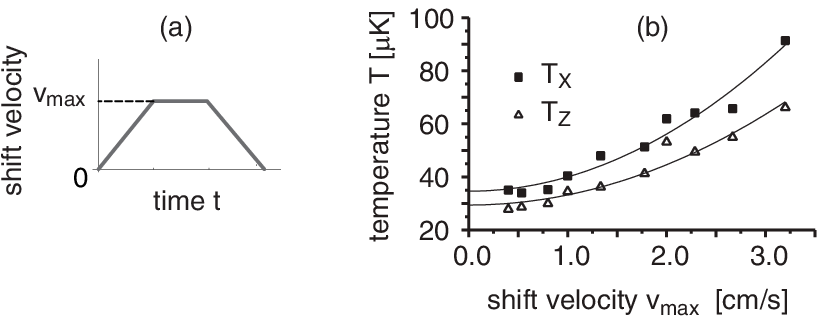}} \caption{(a) The
shift velocity is increased linearly during one spatial period
(0.8\,mm), then kept constant for two periods (1.6\,mm) and ramped
back to zero during the last period. (b) The heating rate
decreases for slow transport. The difference between $T_x$ and
$T_z$ for slow speeds is due to incomplete thermalization during
the preparation stage. For increasing $\vmax$, $T_x$ grows more
rapidly than $T_z$, indicating that heating occurs predominantly
in the transport direction.} \label{fig-motorHeating}
\end{figure}
}

\newcommand{\figF}{
\begin{figure}[\position]
\centerline{\includegraphics{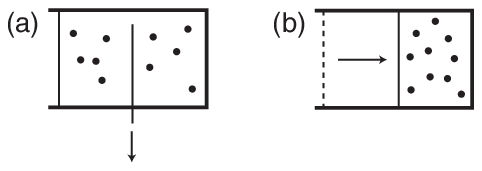}} \caption{Simple
thermodynamic model idealizing the potential of figure
\ref{fig-unite}(a): isolated container divided into halves by a
septum. (a) The septum is pulled out. In the special case where
each half initially contains $N$ atoms at temperature $T_0$, the
temperature, density, and also the phase space density remain
unchanged. (b) The volume is adiabatically reduced by a factor of
2 by slowly moving one of the walls. Again, the phase space
density remains constant. Thus, the complete process doubles the
number of atoms in the right volume without changing the phase
space density.} \label{fig-septum}
\end{figure}
}

\newcommand{\figG}{
\begin{figure}[\position]
\centerline{\includegraphics[scale=0.94]{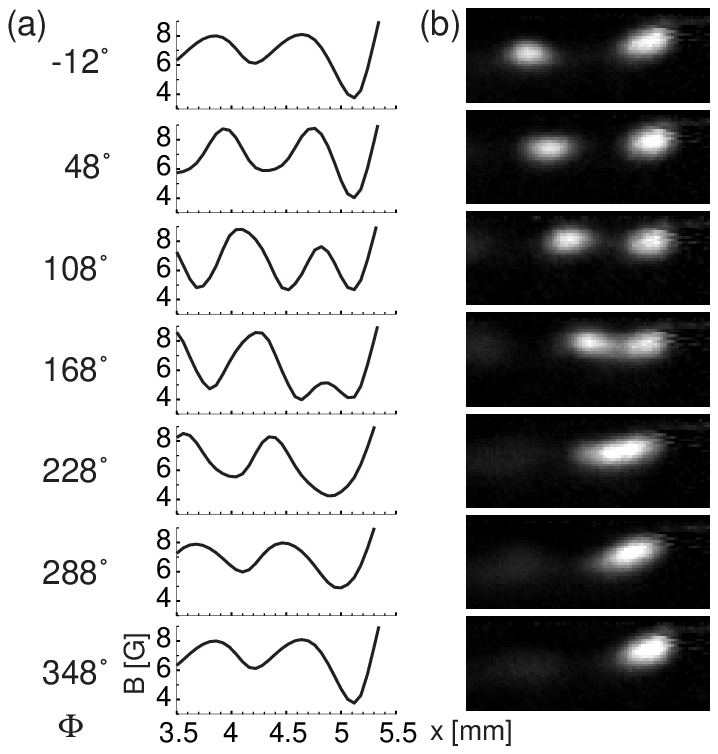}} \caption{(a)
Right part of the magnetic conveyor belt potential when an
additional current $\I{H2}$ is applied (see text). All other
currents and fields are as in fig.~\ref{fig-motorComplete}. During
one cycle of $\Phi$, the potential well arriving from the left
unites with the stationary, right well, which is then compressed
back to its original size while the next well approaches. (b)
Unification of two atom clouds in this potential.}
\label{fig-unite}
\end{figure}
}

\newcommand{\bvec}[1]{{\mathbf{#1}}}

\newcommand{\dip}[1]{\cdot 10^{#1}}

\newcommand{\vmax}{v_{\text{max}}}
\newcommand{\nyosc}{\nu_{\text{osc}}}
\newcommand{\nyR}{\nu_{\text{R}}}

\newcommand{\vecx}{{\mathbf{e}}_x}
\newcommand{\vecy}{{\mathbf{e}}_y}
\newcommand{\vecz}{{\mathbf{e}}_z}

\newcommand{\I}[1]{I_{\text{#1}}}

\newcommand{\vB}[1]{{\mathbf B}_{\text{#1}}}

\newcommand{\Gcmsq}{\mbox{G}/\mbox{cm}^2}

\title{Atom Chip for Transporting and
  Merging Magnetically Trapped Atom Clouds}

\author{W. H\"ansel, J. Reichel, P. Hommelhoff and T. W. H\"ansch}
\address{Max-Planck-Institut f\"ur Quantenoptik and Sektion Physik der
  Ludwig-Maximilians-Universit\"at\\Schellingstr. 4, D-80799
  M\"unchen, Germany }
\date{\today}

\begin{document}

\maketitle

\begin{abstract}
  We demonstrate an integrated magnetic ``atom chip'' which transports cold
  trapped atoms near a surface with very high positioning accuracy.
  Time-dependent currents in a lithographic conductor pattern create a
  moving chain of magnetic potential wells; atoms are transported in
  these wells while remaining confined in all three dimensions. We
  achieve fluxes up to $10^6\,\mbox{s}^{-1}$ with a negligible
  heating rate.  An extension of this ``atomic conveyor belt''
  allows the merging of magnetically trapped atom clouds by
  unification of two Ioffe-Pritchard potentials. Under suitable
  conditions, the clouds merge
  without loss of phase space density. We
  demonstrate this unification process experimentally.
\end{abstract}

\pacs{32.80.Pj, 03.75.-b, 03.67.Lx, 39.90.+d}

Miniaturization makes it possible to design tightly confining
magnetic atom traps with large field gradients and high field
curvature without the need for large currents. With lithographic
or other surface-patterning processes complex ``atom chips'' may
be built which combine many traps, waveguides, and other elements
in order to realize {\it controllable composite quantum systems}
\cite{Zoller00}, as epitomized by the concept of the quantum
computer \cite{foot-QCreview}.

We are experimenting with lithographically produced planar
conducting patterns in an external magnetic bias field as a means
to realize microscopic atom manipulation devices. Although
lithographic neutral-atom traps have been proposed as early as
1995 \cite{Weinstein95}, the difficulty of loading atoms into such
traps has prevented their realization in the past. We have
introduced an efficient loading mechanism \cite{Reichel99} with
the help of a novel mirror-MOT, using a reflecting layer on top of
the circuit pattern to realize the laser fields for laser cooling
and trapping in close proximity to the surface. In this way, we
were able to demonstrate the first lithographic magnetic trap, a
quadrupole trap capturing $^{87}$Rb atoms. More recently, the same
technique has been applied to construct a Ioffe-Pritchard (IP)
trap for $^7$Li atoms \cite{Folman00}.  Integrated parallel
conductors have also been used to realize atom guides in which
atoms are confined in two dimensions and move freely along the
third \cite{Mueller99,Dekker00}. A different, but related approach
was described in \cite{Rosenbusch00}, where a combination of
permanent surface fields and an external bias field produces a
series of parallel magnetic tubes, which propagate as the bias
field vector rotates. In this experiment, too, the mirror-MOT
plays a crucial role in loading the magnetic tubes.

Here we demonstrate an ``atomic conveyor belt'', an integrated
magnetic device proposed in \cite{Reichel99}, which transports
trapped atoms to very precisely controlled positions near a
surface. While the atoms are transported, they remain trapped in
all three dimensions, the rms extension of the trap ground state
being on the order of \mbox{1\,$\mu$m}. The speed of displacement
can be freely controlled by adjusting the frequency of modulating
currents, the device is therefore capable of transporting atoms
adiabatically within well-defined quantum states. As we use a
thermal atomic cloud in the experiment (temperature in the
$30\,\mu$K range), the adiabaticity of the transport is evidenced
by the heating rate, which becomes unmeasurably low for
sufficiently slow transport velocity.

An extension of the conveyor belt allows the merging of atom
clouds trapped in two adjacent Ioffe-Pritchard (IP) traps. The
transfer is complete and, under suitable conditions, occurs
without loss of phase space density. We demonstrate this
unification process experimentally.

Figure \ref{fig-motorLayout} shows the conductor pattern used in
the experiment. It is a slightly modified version of the layout
proposed in \cite{Reichel99}. The width of all relevant conductors
is $50\,\mu$m. In the simplest case, one can create a
quasi-2D-Ioffe-Pritchard trap by applying the current $I_0$ to the
center wire and superimposing the bias field
$\bvec{B}=B_{0,x}\vecx +  B_{0,y}\vecy$. The trap center forms at
distance $r_0 \approx \frac{\mu_0}{2\pi}\frac{I_0}{B_{0,y}}$ from
the surface, and if the distance $r_0$ is much smaller than the
length of the center wire (5.5\,mm in this setup), the
longitudinal potential is box-like with a minimum field strength
of $B_{\text{min}}\approx B_{0,x}$.

For example, by applying a current $I_0=2\,$A ({\it
cf}.~fig.~\ref{fig-motorLayout}) and a bias field
$\vB{0}=0.42\,\mbox{G}\,\vecx+80\,\mbox{G}\,\vecy$, a very
elongated IP trap is formed at a distance of $42\,\mu$m from the
conductor surface. Atom-surface interactions are not expected to
be important at such a large distance \cite{Henkel99b}. For
magnetically trapped $^{87}$Rb atoms this configuration yields
transverse oscillation frequencies of $\nyosc=29$\,kHz in the trap
center, leading to a Lamb-Dicke parameter
$\sqrt{\nyR/\nyosc}=0.36$ with respect to the $^{87}$Rb D2 line
($\nyR=h/(2 m \lambda^2)$ is the recoil frequency). An experiment
with a potential of this type is described in \cite{Reichel00}.

\figA \figB

The conveyor potential as displayed in
fig.~\ref{fig-motorComplete}(a) is formed starting from such a
box-like potential. The field component $B_{0,x}$, which
determines the minimum field strength in the transverse
($\vecy$-$\vecz$) plane, is modulated by means of the currents
$\I{M1}$ and $\I{M2}$. Like a bucket chain, the resulting
potential consists of a chain of trapping wells that can be
continuously displaced along $\vecx$ by adjusting the currents
$\I{M1}$ and $\I{M2}$. For the potential in
fig.~\ref{fig-motorComplete}(a), a constant current $\I{0}=2\,$A
and the bias field $\vB{0}=7\,\mbox{G}\,\vecx+16\,\mbox{G}\,\vecy$
are applied while the modulation currents
\begin{equation}
 (\I{M1},\I{M2}) =1\,\mbox{A}(\cos\Phi,-\sin\Phi)
\end{equation}
create the local minima. As can be seen from figure
\ref{fig-motorComplete}(a), the phase angle $\Phi$ determines the
$x$ position of the minima, so that they can be displaced
continuously over the whole length of the conductor pattern. The
depth of the individual wells, i.~e.\ the magnetic field
difference between the bottom of a well and the saddle point
between two adjacent wells, is 2.5\,G on average and varies
depending on the phase $\Phi$ ({\it
cf.}~fig.~\ref{fig-motorComplete}(a)). Transverse
($\vecy\vecz$-plane) curvatures are between
 $2.5\dip4\,\Gcmsq$ and
 $4\dip5\,\Gcmsq$
near the bottom of the well, leading to oscillation frequencies
between 200\,Hz and 800\,Hz. Longitudinal frequencies are smaller
by a factor of $\sim$\,4.

The experimental setup is an improved version of the one described
in \cite{Reichel99}. Planar gold conductor patterns are formed on
an aluminum nitride substrate using a standard microelectronics
process known as {\it thin-film hybrid technology} (a combination
of photolithography and electroplating). The wire thickness
(height) is $7\,\mu$m and lateral dimensions are down to
$10\,\mu$m. Electrical contacts to the substrate are made by gold
wire bonding.  We load the microtrap from the mirror-MOT
(\cite{Reichel99}, see also \cite{Pfau97}), using a reflecting
silver layer on top of the circuit pattern to realize the laser
fields for laser cooling and trapping in close proximity to the
surface. A thin ($\approx 20\,\mu$m) epoxy intermediate layer
insulates the circuit pattern from the mirror layer and also fills
the gaps between the conductors, leading to a smooth mirror
surface. The MOT is loaded from background vapor (total pressure
in the $10^{-10}$mbar range). It typically produces $4\dip{6}$
cold atoms within a loading time of 5\,s. The atom number is
limited by the small diameter and intensity of our trapping beams
($1/e^2$ diameter of $8.5$\,mm, total power 14\,mW). The
temperature is $30\,\mu$K and the peak density in the
$2\dip{10}\,\mbox{cm}^{-3}$ range after a short optical molasses
phase. The initial MOT employs a ``macroscopic'' quadrupole field
created by external coils and is loaded $\sim$1\,mm away from the
surface. By imbalancing the currents in the coils, the cloud is
shifted towards the surface (``shifted MOT''); the coils are then
switched off and a ``microscopic MOT'' field with the same
orientation is created by switching on $I_0$ and $\I{M2}$ and a
bias field along $\vecy$. This combination yields a chain of
quadrupole-like potentials. Depending on the position of the
shifted MOT, one or several of these small MOTs can be filled with
cold atoms. We use both options, depending on the nature of the
experiment.

\figC

The substrate is mounted upside down in the vacuum cell, so that
the atoms are ``hanging'' below its surface
(fig.~\ref{fig-probeBeam}). Thus, time-of-flight (TOF) images can
be taken of the atomic cloud released from the magnetic trap to
infer its temperature. We observe the atoms by absorptive imaging,
using a probe beam along the $\vecy$ axis. The images show the
horizontal $\vecx$ axis (which is the direction of transport) and
the vertical $\vecz$ axis. The resolution of the imaging system is
$20\,\mu$m.  Sensitivity is sufficient to detect 30 atoms per
pixel without averaging.

The loading procedure ends with an optical pumping pulse which
pumps the atoms to the $F=2,m=2$ ground state. After this pulse,
the magnetic potential is switched on. Here, the initial potential
is the conveyor belt potential
(\textit{cf.}~fig.~\ref{fig-motorComplete}) with the parameter
$\Phi=0$, i.~e., $I_0=2\,$A, $\I{M1}=1\,$A and $\I{M2}=0$. With
this phase value, the positions of the well centers coincide with
those of the quadrupole fields in the previous ``microscopic
MOT'', so that atoms are efficiently transferred. We typically
obtain trapped clouds with $1/\sqrt{e}$ radii $\sigma_x \approx
113\,\mu$m and $\sigma_y\approx 71\,\mu$m, which contain $\sim
1.5\dip{5}$ atoms with a peak density of
$1.5\dip{10}\,\mbox{cm}^{-3}$. The $1/e$ lifetime is about $4\,$s,
limited by the background pressure of rubidium and other residual
gases.

At this point, the conveyor belt is started by varying the phase
$\Phi$. Fig.~\ref{fig-motorComplete}(b) shows absorption images of
the transport process. (As the imaging is destructive, each image
is taken starting from a ``fresh'' trapped sample.) In this first
example, the phase $\Phi$ is made to increase linearly with time,
$\Phi = \omega t$ with $\omega=2\pi/150\,\mbox{ms}$. It should be
emphasized, however, that {\it any} function of time can be used
to control the phase: the atom cloud can be accelerated, or moved
at constant speed, or be stopped at any desired position. The
positioning accuracy is ultimately limited only by the ground
state size, which is $0.9\,\mu$m FWHM for the 800\,Hz oscillation
frequency attained in this potential. The position uncertainty
introduced by current fluctuations is orders of magnitude below
this value.

\figD

In most applications, it is essential to minimize heating of the
trapped cloud during the transport. Leaving aside external
mechanisms such as heating by the surface (which is negligible at
our atom-surface distance of more than 50\,$\mu$m), transitions
may only be induced by the deformation and displacement of the
potential wells during the shifting process.  The rate of such
transitions should decrease to insignificant levels when the speed
of displacement is low enough. In our present experiment, where
many vibrational states are populated, these transitions translate
into heating. We therefore measured the final temperature of the
cloud as a function of the maximum shift velocity $\vmax$
(fig.~\ref{fig-motorHeating}). For decreasing transport speed, the
heating rate decreases, as expected, and becomes unmeasurable
within our $\pm 2\,\mu$K accuracy for speeds $\vmax\le
0.5\,\mbox{cm}/\mbox{s}$.  In this sense, the transport is
adiabatic. When consecutive potential wells are filled with $\sim
1.5\dip5$ atoms each, the shifting process results in a mean flux
of $9.4\dip5$\,s$^{-1}$. In a future experiment where single
vibrational levels are populated, it will be possible to verify
adiabaticity on the quantum mechanical level.

\figE

The above data demonstrate the ability to move and position the
atom cloud. The conveyor belt can be employed in complex
experiments needing precise atom positioning. In cavity QED
experiments, for example, it may be used to transport atoms into
and out of a zone where they interact with a high-finesse optical
resonator, such as a miniature Fabry-P\'erot resonator
\cite{Hood00,Pinkse00} or a silica microsphere
\cite{Treussart94,Vernooy98}. In this way, long and reproducible
interaction times and precise positioning of the atomic sample
within the quantized resonator field can be achieved. Such a
coupling would also enable non-destructive atom detection, as
required for quantum computing. Another, more macroscopic
application of the conveyor belt would be to transport cold atoms
from a production region into a spatially separated interaction
region, providing an integrated version of the coil-based magnetic
transport device which was recently realized in our laboratory
\cite{Greiner00}.

One advantage of the lithographic technique is the ability to
combine different potentials very easily. With only one additional
wire ($\I{H2}$ in fig.~\ref{fig-motorLayout}), a stationary IP
trap can be formed at the end of the conveyor belt, which adds new
options of atom cloud manipulation. The depth of the stationary
trap is controlled by current $\I{H2}$, independently of the
conveyor belt movement. Fig.~\ref{fig-unite}(a) shows the
resulting potential when $\I{H2}(\Phi)$ (directed along $-\vecy$)
is varied according to $ \I{H2}=0.462+0.255\sin(\Phi+0.493)
-0.088\sin(2\Phi-1.482)\,,$ while all other parameters are
controlled as before. (The figure shows only the right part of the
potential, {\it cf}.~fig.~\ref{fig-motorComplete}.) Once per
period the stationary trap above $H_2$ continuously unites with
the conveyor belt trap arriving from the left. The unification
takes place by lowering the barrier between the two traps. At
$\Phi\approx 100^\circ$, the two traps are still fully separated.
At $\Phi\approx 210^\circ$, their separation has vanished, and at
$\Phi\approx 350^\circ$, the united trap has been compressed to
the volume originally occupied by the stationary trap. Thus, the
process is similar to the classical thermostatistical problem of
pulling a septum out of a container, followed by slowly reducing
the container's volume (fig.~\ref{fig-septum}). In particular, if
each trap initially contains $N$ atoms at equal temperatures and
densities, then after one complete cycle, the cloud in the
stationary trap should have the same phase space density as
before. As it now contains $2N$ atoms, temperature must have
increased during the unification.

\figF

\figG

We have implemented this process experimentally by loading two
conveyor traps from the MOT and then applying the current
modulation described above. Fig.~\ref{fig-unite} shows the
computed potentials (a) as well as absorption images (b) of
different stages of the unification process. The potentials have
been designed in a way that equally populated traps should merge
without loss of phase space density. Due to the difficulty of
performing time-of-flight measurements on the two neighbouring
clouds, we have not yet been able to measure the phase-space
density during this process. Instead one can infer the properties
of the unification process by the decrease of phase space density
when only one of the two potential wells is initially populated.
For this situation, we measure a decrease of $\approx 0.6$ per
unification, which is close to the decrease of 0.5 expected for
the unification of two identical traps one of which is empty.

The unification sequence can also be reversed. In this case, an
atom cloud initially localized in the stationary trap is separated
into $n$ parts during $n$ periods. If the separation is done
slowly, the process is adiabatic, opening intriguing possibilities
for the manipulation of atomic wave packets, including
interferometers with trapped atoms. A quantitative study of the
unification and separation processes is currently under way.

This work is supported in part by a grant from the European Union
(ACQUIRE project, contract no. IST--1999--11055).

\end{document}